\documentclass[10pt,letterpaper,amsmath,amssymb,amsfonts]{article}
\usepackage{amsmath}
\usepackage{cite}
\usepackage{graphicx}
\usepackage{dcolumn}
\usepackage{bm}
\usepackage{latexsym,color}
\usepackage{opex3}

\newcommand{\be}{\begin{equation}}
\newcommand{\ee}{\end{equation}}
\newcommand{\ba}{\begin{eqnarray}}
\newcommand{\ea}{\end{eqnarray}}

\begin{document}

\title{Spoof polariton enhanced modal density of states  in planar nanostructured metallic cavities }

\author{P. S. Davids,$^{1*}$ F. Intravaia,$^{2}$ and D.A.R. Dalvit$^3$}

\address{%
$^{1}$Applied Photonics and Microsystems, Sandia National Laboratories, \\ Albuquerque, NM 87185, USA \\
$^2$Max-Born-Institut, 
12489 Berlin, Germany \\
$^3$Theoretical Division, MS B213, Los Alamos National Laboratory, \\Los Alamos, NM 87545, USA 
}%
\email{pdavids@sandia.gov}

\begin{abstract} 
Spoof surface modes on nanostructured metallic surfaces are known to have tailorable dispersion dependent on the geometric characteristics of the periodic pattern. Here we examine the spoof plasmon dispersion on an isolated grating and a grating-planar mirror cavity configuration. The spoof polariton dispersion in the cavity is obtained using the scattering matrix approach, and the related differential modal density of states is introduced to obtain the mode dispersion and classify the cavity polariton modes. The grating-mirror cavity geometry is an example of periodically nanostructured metals above a planar ground plane.  The properties discussed here are relevant for 
applications ranging from thin electromagnetic perfect absorbers to near-field radiative heat transfer.
\end{abstract}



\section{Introduction}

Eigenmodes in planar metal-insulator-metal (MIM) cavities at short separations have been extensively studied owing to the existence of surface plasmon excitations \cite{Castanie2012,Kurokawa2007,Sturman2007}. Surface plasmons are resonant optical excitations at metal-dielectric interfaces and mainly arise due to the material dispersion of the metallic permittivity.  These surface bound optical eigenmodes confine light to extreme sub-wavelength dimensions and many applications require extending surface plasmon confinement and dispersion to the infrared and THz parts of the electromagnetic spectrum \cite{zhang2009slow,kats2011spoof,shen2013ultrathin,sun2013spoof} .
Recently,  Pendry and co-workers \cite{Pendry2004b,Garcia-Vidal2005} have proposed and demonstrated  engineered dispersion by periodically nanostructuring surfaces by perforating perfect electrical conductors.  It has been shown that these perforated nanostructured surfaces support surface modes that have dispersion similiar to real surface plasmons in metals, but with the effective plasma frequency  determined by the geometric parameters of the perforation.   These engineered dispersive surface modes confine light to subwavelength regions and are called spoof surface plasmons. Spoof surface plasmons are also present in nanostructures made of real metals and they have been studied using techniques such as finite difference time-domain \cite{Yu10} and effective medium theory \cite{Rusina10} showing the role of dissipation and dispersion.
In a cavity configuration consisting of a two dimensional periodic frequency selective surface above a metallic ground plane,  these geometrically induced surface modes give rise to perfect infrared absorbers which are actively under development for infrared filtering and detection applications \cite{Peters2010,Peters2012}.  Furthermore, periodically nanostrutured cavities with strong confinement have been examined for THz quantum cascade laser applications \cite{zhang2009slow,kats2011spoof,shen2013ultrathin,sun2013spoof} .

Recently, nanostructured surfaces have been studied within the framework of fluctuation-induced electromagnetic forces and energy transfer \cite{Lambrecht2008,bimonte2009scattering,Davids2010b,DiegoHeatTransfer,Lussange2012}. These setups are cavities formed 
between a planar metallic mirror and a nanostructured grating (or between two nanostructured gratings) separated by vacuum gaps or dielectric material.
Finite temperature equilibrium electromagnetic forces in planar cavities arise due to thermal or vacuum induced fluctuating currents on the cavity mirrors and their associated fluctuating electromagnetic fields \cite{Reid2013}.   By controlling the cavity modal dispersion and density of modes through periodic nanoscale structuring, one can modify the equilibrium cavity interaction \cite{rodriguez2011casimir,Intravaia2013}.
In planar cavities driven out of  thermal equilibrium, such as two plates maintained at different temperatures, nanostructured control of the cavity modes allows for tailored  energy transfer from the hot surface to the cold surface.  At separations shorter than the thermal wavelength, the near-field heat transfer becomes much larger than the black-body contribution \cite{Polder71}  and is dominated by the interaction of the evanescent surface modes existing on the planar surfaces \cite{Carminati1999,Narayanaswamy2008,Narayanaswamy2009}. This enhanced heat transfer has been experimentally measured in a planar cavity configuration \cite{Ottens2011}.
Nanostructuring of these surfaces can be used to tailor and enhance the heat transfer through an specific design of the cavity modes \cite{DiegoHeatTransfer,Lussange2012,Biehs2011}. 

In this paper we focus on the connection of the two previous topics and we study the impact of spoof plasmons on one of the most relevant quantities for the calculation of fluctuation-induced electromagnetic interactions, namely the differential density of states. We consider the case of a cavity formed by grating in front of a metallic plane surface which is the archetype of the geometrical configuration used in most experimental applications. We show how the differential density of states, which describes the difference between the number of modes in the cavity and the number of modes for the isolated surfaces, is strongly affected at low frequencies by the nanostructuring and the presence of spoof plasmons. In section \ref{SpoofModes} of this paper,  we first review the theory of multimodal spoof plasmon dispersion on a perfectly conducting metallic grating, and compare the dispersion for the perfect grating with the computed transverse magnetic (TM) reflection from a finite conductive Au grating.  In section \ref{CavityModes}, the cavity modes in a planar nanostructured cavity are examined. 
We first extend the theory described in section \ref{SpoofModes} to calculate analytically the modes in a cavity formed by a perfectly conducting metallic grating and a perfectly reflecting plane.
Then we study the case where a real metal Au Drude model is used instead of a perfect electric conductor. 
The scattered fields from the metallic planar grating are treated using the Fourier modal scattering matrix method (RCWA) described in  \cite{Davids2010b}. The modal expansion of the transverse fields in the planar cavity is performed and the cavity boundary value problem is treated in the vacuum region between the nanostructured surface and the planar mirror.  
We then introduce and analyze the differential modal density of states, which contains information of the mode structure for both the cavity problem and the isolated surfaces. 
We numerically compute the differential modal density of states as a function of grating-plane separation, and compare it with that corresponding to the plane-plane cavity. This allows us to observe and categorize the spoof plasmons on the nanotructured grating at short separations, as well as the other various modes present in the system.  Finally, section \ref{Conclusions} contains our conclusions.



\section{Spoof surface modes}
\label{SpoofModes}

\begin{figure}[!htb]
\centering \includegraphics[width=8cm]{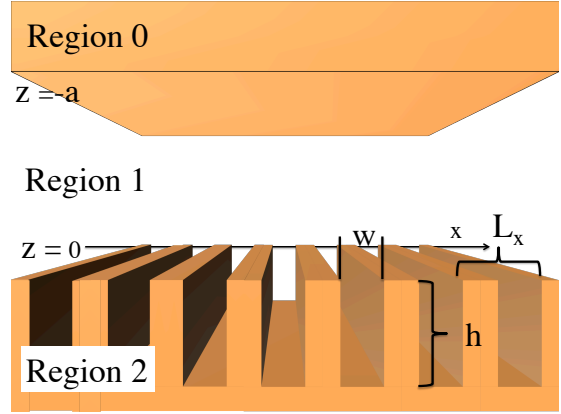} 
\caption{Grating-mirror cavity. The lower mirror is a periodically structured metallic grating.  The upper mirror is a metallic plane at $z=-a$ and the lower mirror top surface is at $z=0$.}
\label{fig:1}
\end{figure}

The simplest structure in which to examine these designer spoof surface plasmon modes is a perfectly conducting grating structure. The lower part of Fig. \ref{fig:1} shows schematically the basic 1D grating.    We begin by considering  the TM surface modes in the perfectly conducting grating structure with period $L_x$,  groove width  $w$, and depth $h$.  The plane of incidence is the $x-z$ plane, therefore $k_y=0$.
The electromagnetic boundary value problem is split into two regions.  In region 1, the vacuum or dielectric region above the etched grating ($z\le 0$), the ${\bf H}$ and ${\bf E}$ field are given by a Bloch mode expansion,
\be
H^{(1)}_y = \sum_n H_n \exp(ik_x^{(n)} x-iq_z^{(n)}  z), \quad E^{(1)}_x = -\sum_n \frac{q_z^{(n)}}{k} H_n \exp(ik_x^{(n)} x-iq_z^{(n)}  z).
\label{fieldsVacuum}
\ee 
Here $k_x^{(n)} = k_x +2\pi n/L_x$ and $q_z^{(n)} =\sqrt{k^2-k_x^{(n)2} }$.  The  ${\bf H}$ field is over the entire period of the grating and satisfies Bloch periodic boundary conditions, and the ${\bf E}$ field is obtained directly from Maxwell's equations.  
Region 2 is defined for $z\ge 0$, the groove is etched in a perfect conductor, thus the electric field is confined in the groove, ($|x|\le w/2$),  and must vanish at the bottom of the groove. If we consider only the fundamental mode,  the transverse magnetic field and electric field in the groove the electric field are given by
\be
H^{(2)}_y = -iB\cos(k(z-h)),  \quad E^{(2)}_x = B\sin(k(z-h)),
\ee
and $E^{(2)}_z = 0$.  
The boundary conditions at the grating top surface, $z=0$, require the continuity of the tangential ${\bf H}$ fields.  By integrating over the groove width we obtain 
\be
 \sum_n H_n s_n = -iB \cos(kh), \quad s_n =\frac{\sin\left( k_x^{(n)}w/2\right)}{ k_x^{(n)}w/2 }.
\label{spp1}
\ee
The continuity of the tangential electric field at $z=0$  and using the orthogonality of the Fourier mode expansion, we obtain 
\be
H_n = B \frac{k}{q_z^{(n)}}\frac{w}{L_x} \sin(kh) s_n.
\label{spp2}
\ee
Combining Eq. (\ref{spp1}) and Eq.(\ref{spp2}), we obtain the resonance condition for the spoof surface plasmons,
\be
\frac{kw}{L_x} \sum_n \frac{s_n^2}{\sqrt{k_x^{(n)2} - k^2}} = \cot(kh),
\label{spp:eq1}
\ee
for the multimodal single grating \cite{Jiang2009}.
For a single mode ($n=0$),  the spoof surface mode is 
\be
\frac{w}{L_x} s^2_0 \tan(kh) = \frac{\sqrt{k_x^{2} - k^2}}{k},
\ee
which reduces to the standard result \cite{Pendry2004b,Garcia-Vidal2005}.

\begin{figure}[b]
\vskip -0.85cm
\centering \includegraphics[width=12.5cm]{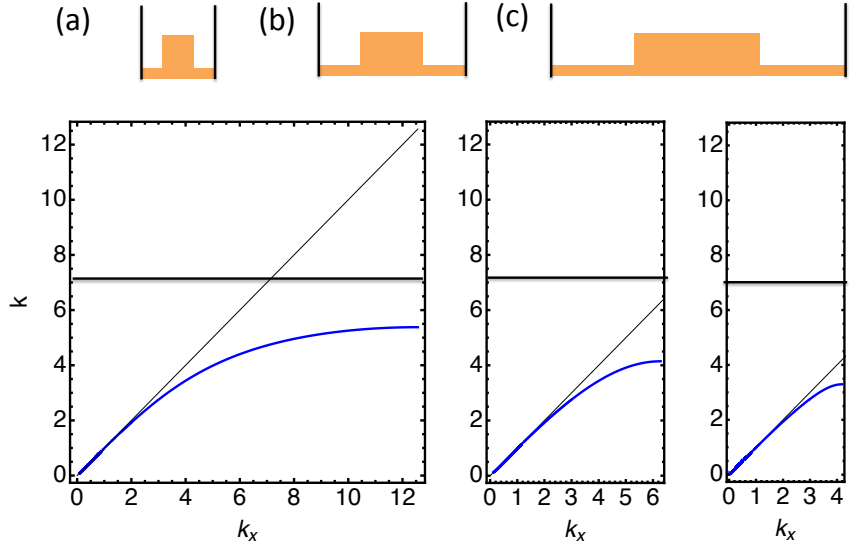} 
\vskip -1.0cm
\caption{Spoof plasmon dispersion as a function of the grating period for fixed duty cycle and groove depth.  The duty cycle is 64\% and the grating depth is $h =216$ nm for all cases.  (a)  $L_x =250$ nm and $w=160$ nm; (b) $L_x=500$ nm and $w=320$ nm; (c)  $L_x=750$ nm and $w=480$ nm. The horizontal line indicates the effective plasma wave-vector $k_{pl} =\pi/2h$. The units of the $k$ and $k_x$ are both $\mu \text{m} ^{-1}$. }
\label{fig:2}
\end{figure}

Eq. (\ref{spp:eq1})  is a transcendental equation for the general dispersion of a resonant excitation of the perfect metallic grating.  The solution of the transcendental equation gives the dispersion relationship for spoof plasmons.
Figure \ref{fig:2} shows the computed spoof plasmon dispersion relationship from a perfect conducting grating with the geometric dimensions outlined in the caption.   In Fig. \ref{fig:2},  the grating duty cycle and depth are held fixed while the grating period is allowed to vary.
 The effective spoof surface plasmon wavelength for these geometric parameters can be obtained from the asymptotic $k_x$ behavior.  The same effective plasmon wavelength of  $\lambda_{pl} \approx 864$ nm  is obtained for the grating parameters in the caption. From Fig. \ref{fig:2},  it appears that the spoof mode dispersion is only weakly dependent on the period with fixed duty cycle. The main impact of the period is to reduce the Brillouin zone.   The effective spoof plasmon wavelength should be contrasted with plasma wavelengths for real metals, which occur in the ultraviolet portion of the spectrum.  For instance,  the Au plasma wavelength used in our Drude model is $\lambda_{pl} = 146$ nm \cite{Pendry2004b,Garcia-Vidal2005}. The nature of the dispersive mode branch below the light line strongly confines the  mode to the surface due to the exponential decay of the Bloch mode. The effect of nanostructuring on the perfect conductor has created dispersive surface modes which mimic surface plasmons with much lower equivalent plasma frequencies.  
\begin{figure}
\centering \includegraphics[width=12.5cm]{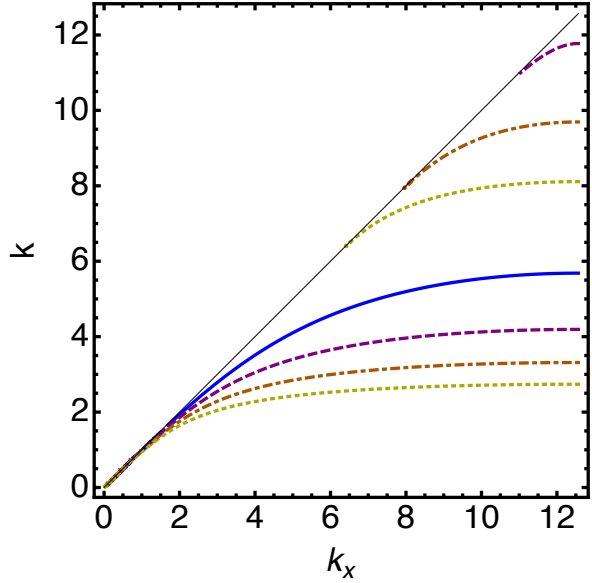} 
\vskip -0.5cm
\caption{Spoof plasmon dispersion relation as a function of the groove depth $h$. The grating period  and width are  fixed at $250$ nm and $160$ nm, respectively. The groove depth is $h=200$ nm (blue solid line),  $h=300$ nm (purple dashed lines), $h=400$ nm (red dashed lines), and  $h=500$ nm (gold dashed lines). As we increase the value of $h$ we observe the appearance of high order modes above the fundamental, indicating the relevance of the structure's depth for the dispersion relation.}
\label{fig:3}
\end{figure}

Figure \ref{fig:3} shows the influence of the grating depth on the geometrically induced surface mode for fixed period and duty cycle.   Here the depth of grating is varied from $200$ nm to $500$ nm. The induced surface mode dispersion is seen to be very sensitive to grating depth, and becomes multimodal at grating depths larger that $200$ nm for the chosen period and duty cycle.  All the high-order modes have a cutoff at the lightline for the value $k=k_{x}=m \pi / h$ \cite{Jiang2009}. Indeed, the first term of the series in \eqref{spp:eq1} diverges for $k=k_{x}$ which leads to the solutions of $\tan(k h)=0$.
This strong dependence of the spoof plasmon dispersion on the grating depth makes the system easily tunable for various applications. 

\begin{figure}
\vskip -0.25cm
\centering \includegraphics[width=13.5cm]{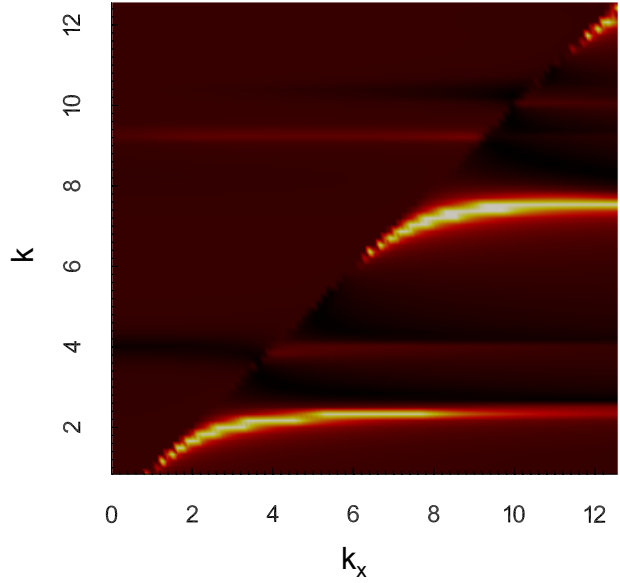} 
\vskip -0.75cm
\caption{Zeroth order TM polarized reflection at normal incidence from a grating with finite conductivity given by the Drude model parameters for Au.  The geometrical parameters of the grating are $h=400$ nm, $L_x=250$ nm, and $w=160$ nm.
The bright features below the light-line ($k<k_x$) represent poles of the reflection matrix.
}
\label{fig:4}
\end{figure}

The geometrically induced surface plasmon modes are strictly speaking found on the perfect conducting grating.  It is natural to inquire if the spoof plasmons occur on a metallic grating with finite conductivity.  Figure \ref{fig:4} shows the computed zeroth order reflection for TM polarized incident light using the Fourier modal scattering method (RCWA)  from a Au grating (see also section \ref{MetallicCavity}). We use a Drude model permittivity for gold  given by
\be
\epsilon(\omega) = 1- \frac{\omega^2_{pl}}{\omega (\omega + i \gamma)},
\ee
where $\omega_{pl} =1.27524 \times10^{16} ~\text{sec}^{-1}$ and $\gamma  =6.59631 \times10^{13} ~\text{sec}^{-1}$.  
There is qualitative agreement between the numerically computed modal dispersion for the finite conductive case with the analytic dispersion for the perfect conducting grating shown in Fig. \ref{fig:3} for the groove depths between  $400$ to 500 nm. The finite conductivity and the dispersion of the grating's permittivity affect the values of the effective spoof plasmon dispersive resonances by shifting and increasing the resonances' width and by effectively modifying the grating's geometrical parameters through a finite penetration depth $\delta_{\omega}$ (in this case $\delta_{\omega}\sim 25 nm$ for $k=\omega/c\sim 10$ $\mu$m$^{-1}$).


\section{ Cavity Modes}
\label{CavityModes}

In this section, we examine  electromagnetic resonances in a planar periodically nanostructured cavity represented in Fig. \ref{fig:1}. 
The cavity consists of a planar mirror (surface 1) located  at $z=-a$, and the periodically modulated grating substrate (surface 2) lies at $z=0$.   Explicitly, we define region 0 as a semi-infinite mirror region corresponding to surface 1, i.e. $-\infty < z \le -a$; region 1 is the cavity vacuum gap between the planar mirror and the planar top surface of the grating, i.e.  $-a\le z \le 0$; and region 2 is the etched grating region, i.e.  $z\ge 0$.  
The cavity's modal properties  are obtained by describing the fields in these three regions, and by requiring the continuity of the transverse fields across each of the planar interfaces \cite{Davids2010b,Intravaia2013}. In the following we are going to analyze this geometry in two cases. In the first case, both the mirror and the grating are made of a perfect electric conductor. As in the previous section most of the calculations can be performed analytically. The results will then be compared with the second case, where the cavity is formed by a plane metallic mirror above  a metallic grating described by the Au Drude model.


\subsection{Perfect electric conductors}

In our first scenario, the use of a perfect electric conductor greatly simplifies the application of the boundary conditions. Indeed, with respect to the procedure described in the previous section, the presence of the plane at $z=-a$ only amounts to replacing the fields in \eqref{fieldsVacuum} with the following expressions
\be
H^{(1)}_y = \sum_n H_n \exp(ik_x^{(n)} x)\cos(q_z^{(n)} (z+a)), \quad E^{(1)}_x = i\sum_n \frac{q_z^{(n)}}{k} H_n \exp(ik_x^{(n)} x)\sin(q_z^{(n)}  (z+a)).
\label{fieldsVacuum}
\ee  
Following the same reasoning as above one obtains
\begin{equation}
 -k\frac{w}{L_x}\sum_n   \frac{s^{2}_n}{q_z^{(n)}}  \cot(q_z^{(n)} a)=  \cot(kh),
\label{eq:spp1}
\end{equation}
which in the limit $a\to \infty$ reduces to the result of the previous section.

\begin{figure*}[hhh]
\centering
 \includegraphics[width=6.4cm]{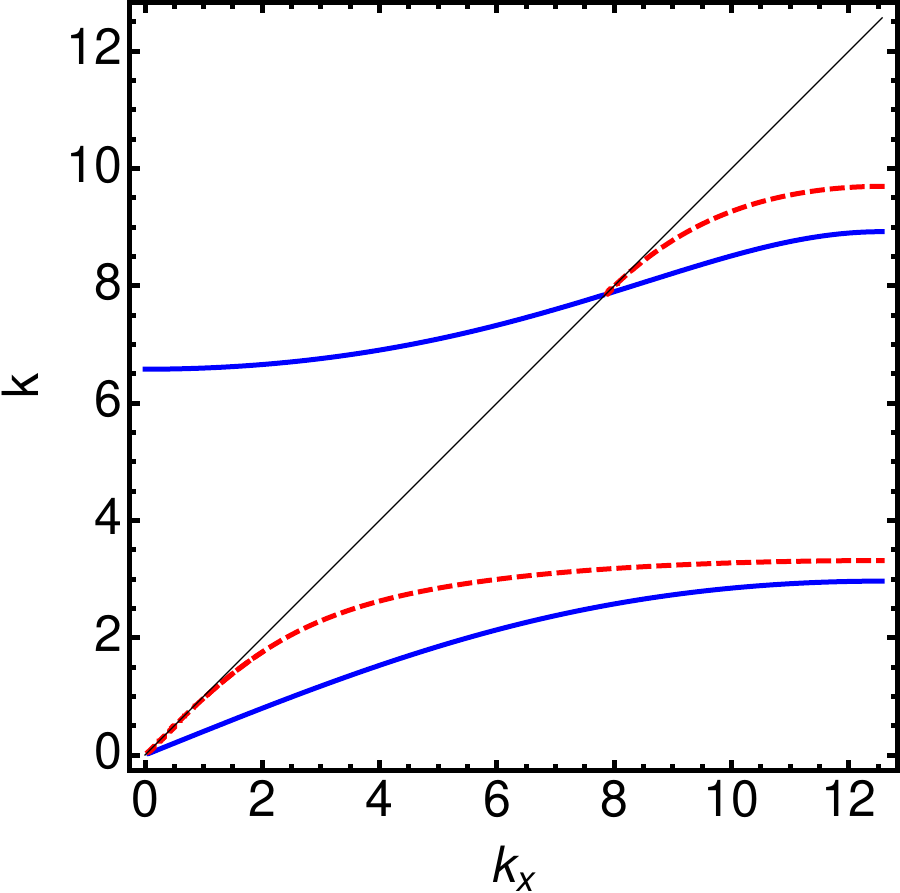}
 \hspace{0.2cm}
\includegraphics[width=6.4cm]{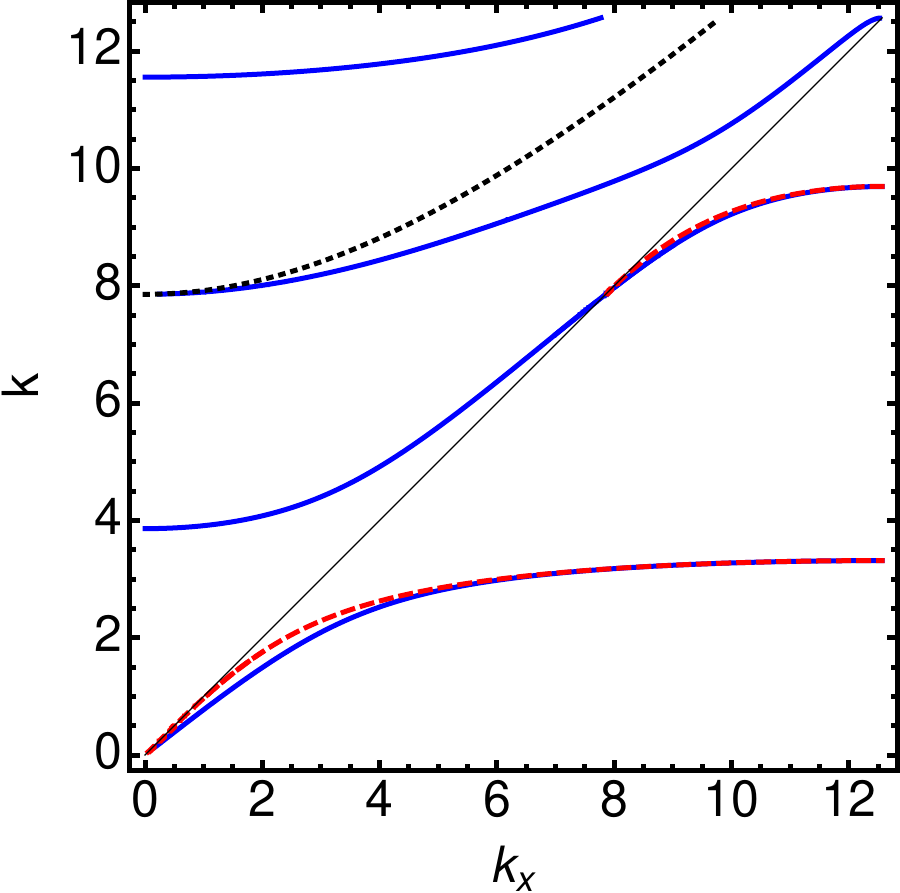}
\caption{Modes in a perfectly conducting grating-plane cavity (full blue lines) compared with the spoof-plasmons for the isolated grating (red dashed lines). The grating parameters are $L_x =250$ nm, $h=400$ and $w=160$ nm and the cavity length is $a=50$ nm (left) and $a=400$ nm (right). The thin black line is the lightline separating the propagating (above the line) from the evanenscent sector (below the line). The dotted black line in the right plot indicates the modes for a plane-plane cavity of the same separation. The units of $k$ and $k_x$ are both $\mu \text{m} ^{-1}$. }
\label{cavity_disp_perf}
\end{figure*}

Figure \ref{cavity_disp_perf} shows the computed cavity mode dispersion relationship for the geometric dimensions outlined in the caption. At short distances the cavity modes are red-shifted spoof-plasmons resonances (as comparison we also plot the grating's spoof-plasmons). Note that because of the planar mirror,  the higher order spoof-plasmon modes are now allowed to cross into the propagating sector, intersecting the lightline at $k=k_{x}=m\pi/h$. At larger separations pure propagating modes start to appear, and in the evanescent region the cavity spoof-plasmons quickly tend towards the modes for the isolated grating.


\subsection{Metallic case: modes and differential density of states}
\label{MetallicCavity}
We now consider the case where the cavity grating and the mirror are no longer perfect conductors, but are made of finite conductive metal described by a Drude mode for the Au grating and mirror. The symmetry and the underlying periodicity of our system implies that the modes are naturally expanded in a Bloch planewave mode basis, leading to  
\begin{eqnarray} 
\Psi_0 &  = & \sum_{\mu} C_{\mu} X^{(-)}_{\mu,0} e^{-iq_z^{(0)}z} \label{cav1}\\
\Psi_1 &  = & \sum_{\mu}  A_{\mu} X^{(+)}_{\mu,1} e^{iq_z^{(1)}z} +B_{\mu} X^{(-)}_{\mu,1} e^{-iq_z^{(1)}z} \label{cav2} \\
\Psi_2 &  = & \sum_{\mu} D_{\mu} X^{(+)}_{\mu,2} e^{iq_z^{(2)}z}, \label{cav3}
\end{eqnarray}
where $\Psi$ denotes either the ${\bf E}$ or ${\bf H}$ field,  $q^{(i)}_z = \sqrt{k^2\epsilon_i - q_n^2-q_m^2}$ is the propagation eigenvalue for the uniform media in the $i$-th region, and $\mu = (n,m,\sigma)$ where $\sigma=s,p$ denotes the state of polarization. Here $q_n=k_x + 2\pi n/L_x$ and  $q_m=k_y + 2\pi m/L_x$ ($n,m$ are integers). 
The boundary conditions on the continuity of the transverse fields at the two interfaces ($\Psi_0(z=-a) = \Psi_1(z=-a)$ and $\Psi_1(z=0)= \Psi_2(z=0)$) allow for the determination of the coefficients $A_{\mu}$, $B_{\mu}$ $C_{\mu}$ and $D_{\mu}$. 

In previous work, the Fourier modal solution for scattering from the periodically modulated surface using both S-matrix and T-matrix techniques has been described \cite{Davids2010b,intravaia2012quasianalytical,Li1996,Li1996a,Li1997,Li2003,Li2003c}.  Here, we will use the S-matrix formalism to obtain a secular equation for the eigenmodes in the cavity. The boundary condition at the grating  top surface  ($z=0$)  is solved by  the exact reflection matrix, $R$, which can been determined for arbitrary input amplitude $A_{\mu}$. Eq. (\ref{cav2}) can be rewritten as
\be
\Psi_1   =  \sum_{\mu} A_{\nu} \left(  \delta_{\mu,\nu}X^{(+)}_{\mu,1} e^{iq_z^{(1)}z} +\sum_{\nu}  R_{\mu,\nu}  X^{(-)}_{\mu,1} e^{-iq_z^{(1)}z}\right),
\ee
where $B =  R A$ in matrix notation. The use of the  S-matrix implies that we have satisfied the continuity  of the transverse field and scattering boundary conditions at $z=0$.  The boundary conditions at the planar mirror at $z=-a$ are the same as for the planar cavity case, and we obtain
\begin{equation}
C_{\mu} e^{iq^0_z a} = i\alpha_{\mu} A_{\mu} e^{-iq_z^1 a} + \beta_{\mu} B_{\mu} e^{iq_z^1 a},
\end{equation}
where 
\begin{equation*}
\alpha_{\mu}=
\begin{cases}
\frac{1}{2\sqrt{q_z^{(0)}q_z^{(1)}}} (q_z^{(1)}-q_z^{(0)}) & \sigma=s  \\
\frac{1}{2\sqrt{q_z^{(0)}q_z^{(1)}}}\left(\sqrt{\frac{\epsilon_0}{\epsilon_1}} q_z^{(1)}-\sqrt{\frac{\epsilon_1}{\epsilon_0}}q_z^{(0)} \right)  & \sigma=p,
\end{cases}
\end{equation*}
and 
\begin{equation*}
\beta_{\mu}=
\begin{cases}
\frac{1}{2\sqrt{q_z^{(0)}q_z^{(1)}}}(q_z^{(1)}+q_z^{(0)})  &\sigma=s \\
\frac{1}{2\sqrt{q_z^{(0)}q_z^{(1)}}}\left(\sqrt{\frac{\epsilon_0}{\epsilon_1}} q_z^{(1)}+\sqrt{\frac{\epsilon_1}{\epsilon_0}}q_z^{(0)} \right)  & \sigma=p.
\end{cases}
\end{equation*}
The $z$ propagation terms, $q_z$,  depend on the discrete spatial frequency index, $\mu = (nm,\sigma)$ and the polarization index  $\sigma = $s or p. The projection of the cavity mode onto the planar mirror mode is given by
\begin{equation}
C_{\mu} = -\frac{i}{\alpha_{\mu}} A_{\mu} e^{-i (q_z^1+q_z^0) a}. 
\end{equation}
This leads to the following secular equation 
\begin{equation}
\sum_{\nu} \left( -i\delta_{\mu,\nu} \left(\frac{1}{\alpha_{\mu}} + \alpha_{\mu} \right)e^{-iq_z^1a} -e^{iq_z^1a} \beta_{\mu} R_{\mu,\nu} \right) A_{\nu} = 0
\end{equation}
which can be simplified to give
\begin{equation}
\sum_{\nu} \left( \delta_{\mu,\nu}  - e^{2iq^{(1)}_{z,\mu}a} \rho_{\mu} R_{\mu,\nu} \right) A_{\nu} = 0.
 \label{secular_eq}
\end{equation}
The cavity scattering matrix is defined as $\mathbf{D}_{\mu,\nu} =  \delta_{\mu,\nu}  - e^{2iq^{(1)}_{z,\mu}a} \rho_{\mu} R_{\mu,\nu} $,  where 
\begin{equation}
\rho_{\mu}=
\begin{cases}
ir_s= i \left(\frac{q_z^{(1)} -q_z^{(0)}}{ q_z^{(1)} +q_z^{(0)}}  \right)  & \sigma=s \\
ir_p = i  \left(\frac{  \epsilon_0 q_z^{(1)} -\epsilon_1 q_z^{(0)}}{ \epsilon_0 q_z^{(1)} + \epsilon_1 q_z^{(0)}}  \right) & \sigma=p,
\end{cases}
\end{equation}
is the diagonal reflection matrix,  and $r_s$ and $r_p$ are the  Fresnel reflection coefficients from the planar interface.   The generalized secular equation for the periodically modulated mirror cavity is now a matrix equation
\begin{equation}
\det( {\bf I} -e^{iq_za} \cdot \boldsymbol{\rho} \cdot e^{iq_za}\cdot \mathbf{R} ) = 0 .
\label{secular_gen}
\end{equation}
Here  $e^{iq_za}$  are the diagonal propagation matrices.
The solution to the general secular equation (\ref{secular_gen}) for $(\omega, k_x,k_y)$ represents the dispersion relationship for the cavity resonances which depend parametrically on the mirror separation distance, $a$.  

The cavity polariton dispersion is then obtained from the zeros of the secular determinant of the cavity scattering matrix, which are  in general  complex valued because of the dissipation in the metal.  The real part of the complex zero represents the cavity resonance frequencies, $\omega_n(a)$,  and the  imaginary frequency component the resonance linewidth.  Rather than solving for the  dispersion directly, it is convenient to examine the analytic properties of the secular determinant  and the modal dispersion relationship by constructing the  differential modal density of states.   The differential modal density of states is related to the secular determinant by
\begin{equation}
\Delta n(\omega,k_x,k_y,a)=-\frac{1}{\pi}\mathrm{Im}\partial_{\omega}\log\mathrm{det}(\mathbf{I}-e^{i q_{z} a}\cdot \boldsymbol{\rho}\cdot e^{i q_{z} a}\cdot \mathbf{R}),
\label{DiffDOS1}
\end{equation}
where $\Delta n= n(\omega,k_x,k_y,a)-n(\omega,k_x,k_y,\infty)$, and $n(\omega,k_x,k_y,a)$ gives the number of modes per unit of frequency for a cavity separation $a$. In this respect $n(\omega,k_x,k_y,\infty)$ represents the density of states for an infinitely separated cavity (isolated surrfaces).    Eq. (\ref{DiffDOS1}) arises from the decomposition of the  secular determinant into the ratio of determinants of  transfer matrices which is used to isolate the zeros and poles of the cavity scattering matrix (see \cite{intravaia2012quasianalytical} for details).  This results in the well known  Krein formula  for the differential density of states and is related to a contour integral in the complex frequency plane \cite{wirzba2008casimir}.  Here the contour is restricted to the real positive frequency axis  and  Eq. ({\ref{DiffDOS1}) is a function of a real frequency variable.  
In Eq. (\ref{DiffDOS1}), the secular determinant may possess singularities that arise from poles in the  planar or grating reflection matrices.  These poles  correspond to electromagnetic resonances  on the isolated surfaces and contribute to the differential density of states through $n(\omega,k_x,k_y,\infty)$.  The poles of $\rho$  correspond to isolated surface plasmon modes on  the planar metal surface, and the  poles of $\mathbf{R}$ correspond to the spoof surface plasmon modes on the isolated grating surface.   We will examine the differential modal density of states and infer the cavity mode dispersion within this context.  

\begin{figure*}[!htbp]
\centering \includegraphics[width=11.5cm]{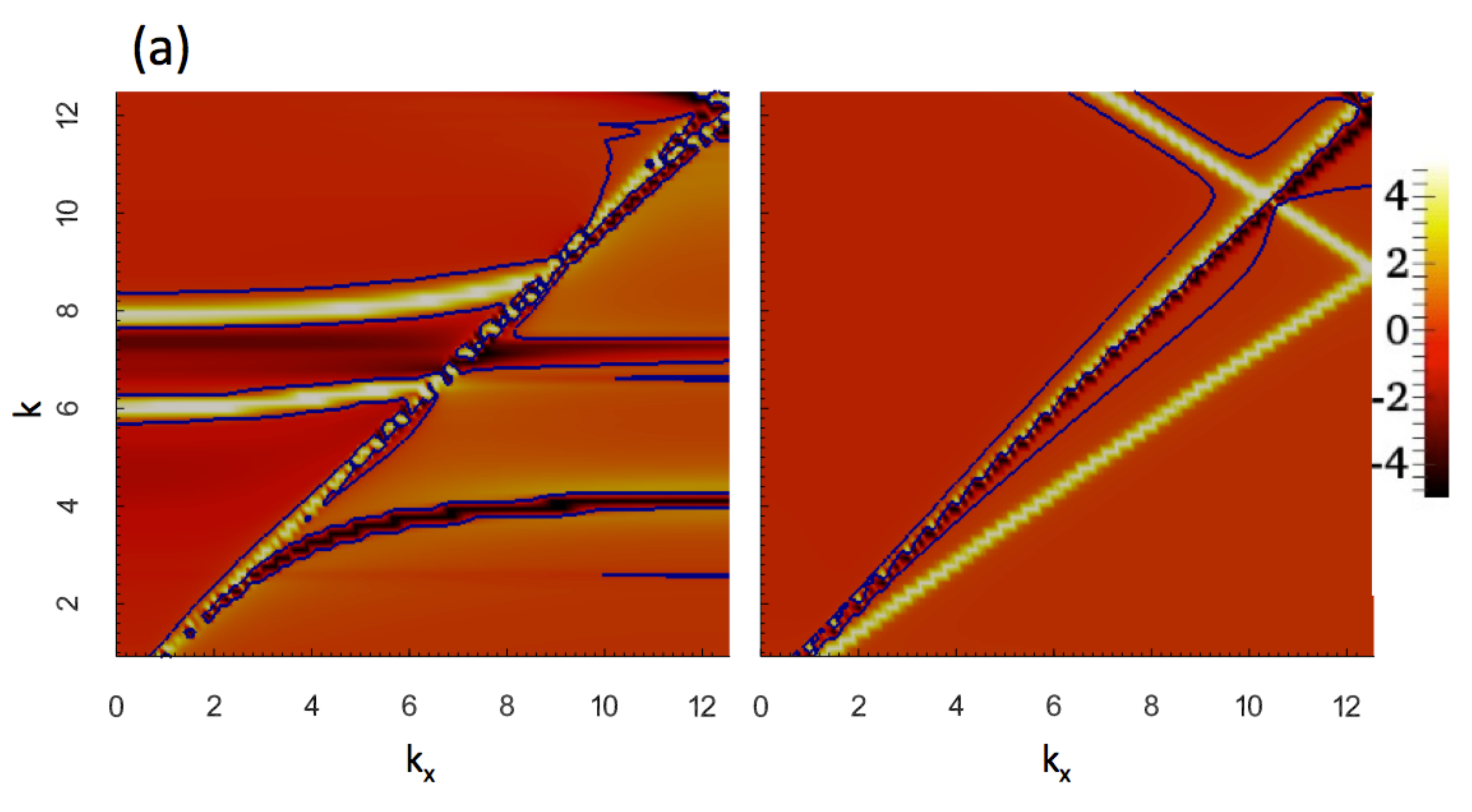} \\
\centering \includegraphics[width=11.5cm]{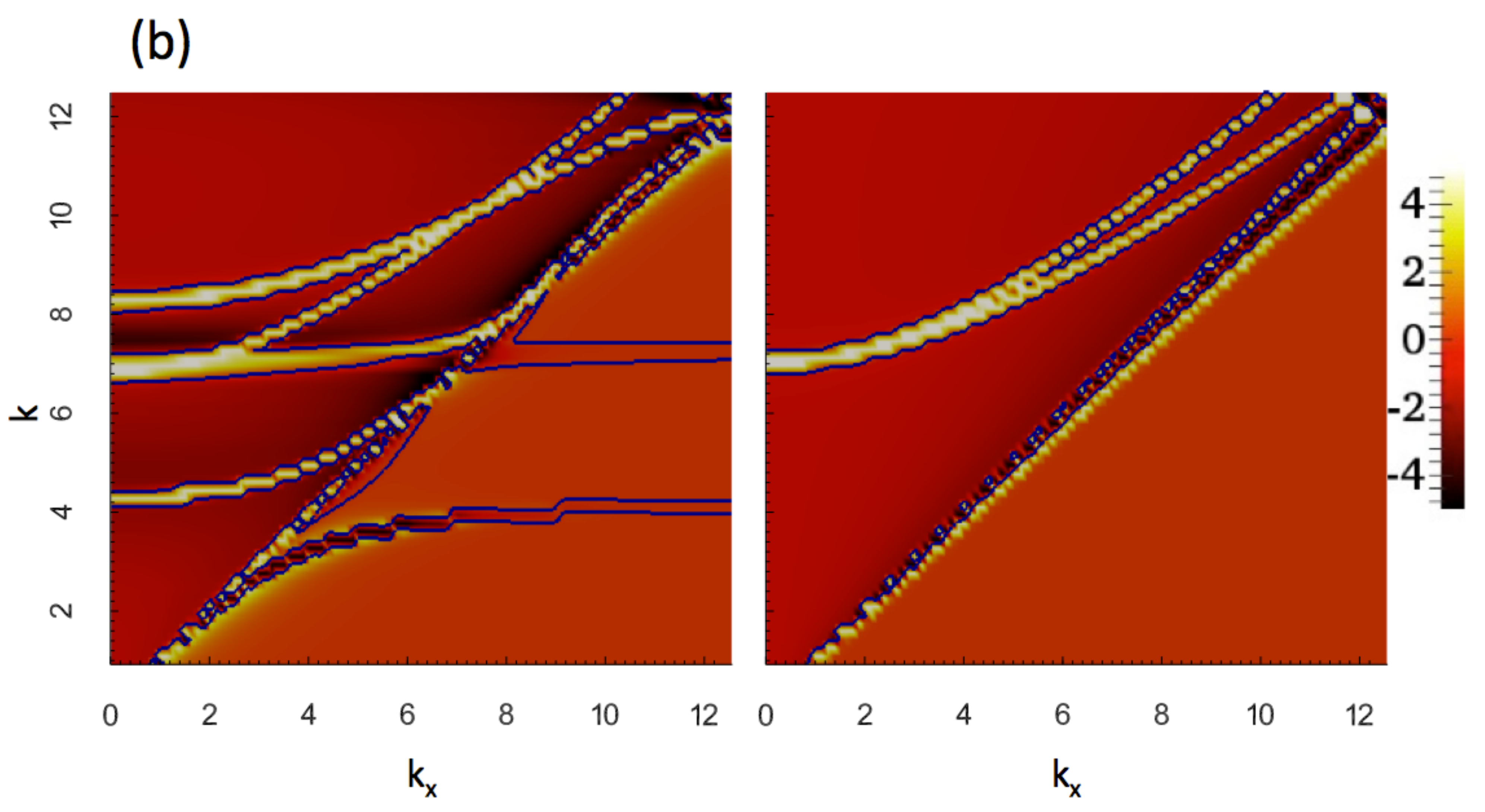} \\
\centering \includegraphics[width=11.5cm]{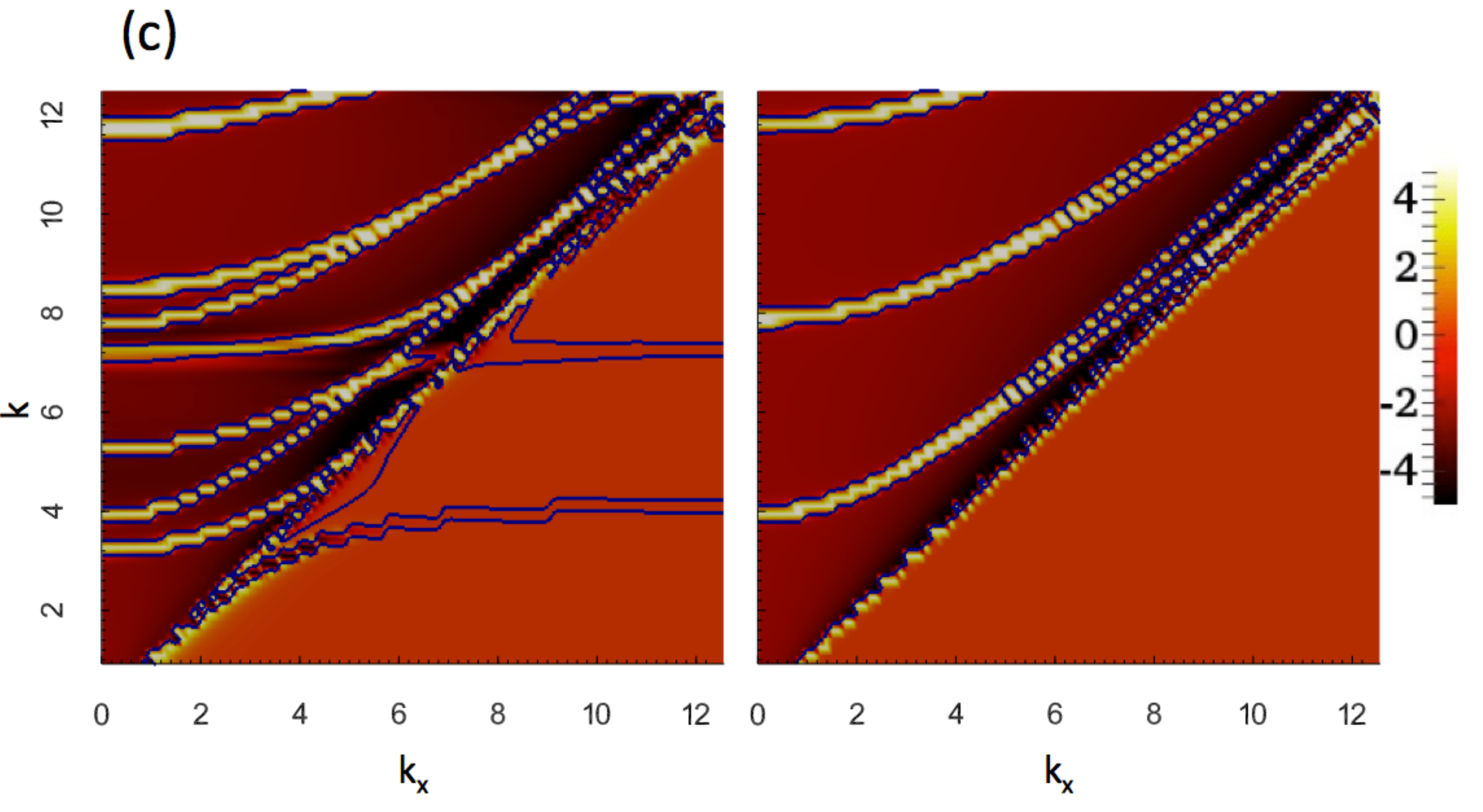}  \\
\caption{Differential modal density of states  of a nanostructured grating-plane cavity (left panels) compared to plots of a plane-plane cavity, for various separations 
(a) $a=50$ nm; (b) $a=400$ nm; (c) $a=750$ nm.  In all these plots $k_y$ is fixed to zero. The zero density contours $\Delta n=0$ are highlighted in blue.
The parameters of the grating are $L_x=250$ nm, $w=160$ nm, and $h= 216$ nm (these parameters correspond to the metallic grating studied in recent Casimir force measurements \cite{Intravaia2013}). Both the grating and the plane are described using the Drude model for gold.
The units of $k$ and $k_x$ are both $\mu \text{m} ^{-1}$, while the density of states is in units of $\pi c$/sec. }
\label{cavity_disp}
\end{figure*}


\subsection{Spoof plasmons in a metallic cavity}

The cavity differential modal density of states and dispersion is examined graphically for different plane grating spacings. 
The differential density of states is computed numerically from the complex secular determinant and we plot Eq. (\ref{DiffDOS1}) as a function of $k=\omega/c$ versus $k_x$, fixing $k_y=0$. Figure \ref{cavity_disp} shows  the computed differential modal density of states, and  illustrates the mode  dispersion of the grating-plane cavity (left panels) and of the plane-plane cavity (right panels).  The series (a)-(c) shows the cavity dispersion comparison as a function of separation.  The dispersion is naturally separated into regions below the light-line $k<k_x$, and regions above the light-line $k> k_x$.   Furthermore, the sign of the differential modal density of states provides insight into the  type of resonance.  
Positive values of Eq. (\ref{DiffDOS1}) (bright regions) correspond to positive differential modal density of states and represent resonant cavity modes at finite separation.  Negative graphical values (dark regions) correspond to negative differential density of states and come from the resonant modes on the isolated surfaces or the continuum of propageting modes in the cavity \cite{intravaia2012quasianalytical}. 
In the following, we will examine metallic cavities with a vacuum gap that are modeled using the Au Drude model described in previous sections.  

For the grating-plane cavity (left panels) of Fig. \ref{cavity_disp} the differential density of states is plotted with the $\Delta n = 0$ contours shown in blue.  Below the light-line, ($k_x > k$), at the shortest separation,  we find a dark line region that is bracketed by the $\Delta n = 0$ contours and a bright region  corresponding to positive differential density of states is seen above the dark line.  The corresponding dispersive modes for the dark region are the $p$-polarized spoof polariton that occurs on the isolated Au grating (see also Fig. \ref{fig:4}) and the bright region above correspond to $p$-polarized spoof plasmon cavity resonant mode, respectively.  They can be compared to the analogous PEC spoof polariton modes in Fig. \ref{cavity_disp_perf}.  At small grating mirror separations,  the differential modal density of states is substantially different from zero because the cavity and the isolated modes are different.  As the separation is increased,  these modes start to overlap and the differential density of states is reduced (see also Fig. \ref{cavity_disp_perf}).  This happens first at large $k_x$ due to the Rayleigh propagation factors in Eq.\eqref{DiffDOS1},  which are  exponentially decaying below the light-line.
Above the light-line, bright line propagating cavity modes are evident.  The  splitting of these cavity modes is due to $s$ and $p$ polarized reflections from the grating, which are very different for the two polarizations. At larger separations, we see that above the light-line  the modal density of states has bright line modes that occur on a negative background that corresponds to the continuum modes in the cavity.  The bright line cavity modes refer to a positive differential modal density of states   and also exhibit avoided crossing behavior  in the multimodal cavity.
 
For the  planar cavity  (right panels) in Fig. \ref{cavity_disp},  we see bright-line parabolic mode dispersion above the light-line on the continuum (dark) background.  For small $k_x$, there is negligible brigth-line splitting since the reflection from a planar mirror is degenerate for $s$ and $p$ polarization.   This is in sharp contrast to the large polarization mode splitting seen for the grating cavity.  For larger $k_x$ near the light line, the planar parabolic cavity modes are weakly split due to small differences in large angle $s$ and $p$ polarized reflections from planar surfaces.   Below the light-line, we can see both a dark line and a bright line mode near the light line.  The bright line mode is symmetric red-shifted metal-vacuum-metal surface plasmon dispersion mode which  asymptotically approaches the plasma frequency for Au, $k_{pl}=2\pi/\lambda_{pl}$ with $\lambda_{pl} = 146$ nm from below \cite{Intravaia07} (the antisymmetric mode is out of the plotted range). The dark line mode corresponds to isolated surface plasmon modes on the individual metal surfaces.  At large separations, the bright line approaches the isolated mode and they become degenerate.   
An accurate inspection of the right left panel show that modes similar to the one just described also appear in the plane-grating geometry. These modes correspond to the isolated plasmon living on the mirror plane and the corresponding cavity plasmon with a difference frequency due the interaction with the grating.

\begin{figure}[!htbp]
\centering \includegraphics[width=13.5cm]{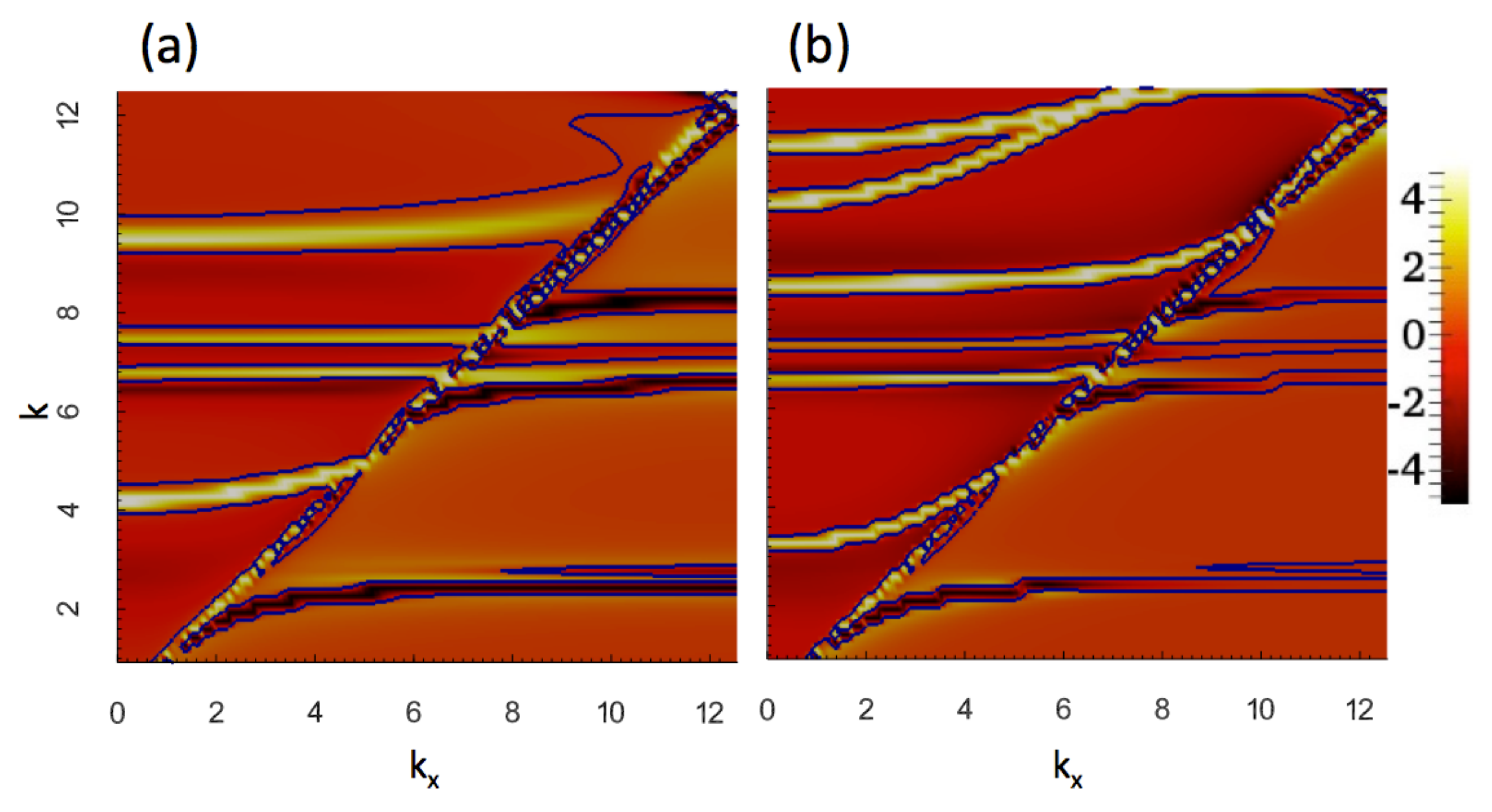} 
\caption{Differential modal density of states plots for nanostructure grating-plane cavity for different separations: (a) $a=50$ nm  and (b) $a=250$ nm.  The period and the width are the same as in fig. (\ref{cavity_disp}), but the depth is different, $h = 400$ nm. The units of $k$ and $k_x$ are both $\mu \text{m} ^{-1}$, while the density of states is in units of $\pi c$/sec. }
\label{fig6}
\end{figure}

Figure \ref{fig6} shows the computed cavity differential modal density of states and mode dispersion for the same grating as in figure \ref{cavity_disp} but more deeply etched,  $h=400$ nm. Multimodal cavity surface polaritons are seen below the light line at very short mirror grating separation, $a=50$  to $250$ nm.  This result is consistent with the prediction of multimodal spoof plasmon grating results shown in Fig. \ref{fig:3} and  Fig. \ref{cavity_disp_perf} for the perfect conducting single grating and grating-plane cavity.   Variation of the grating depth is seen to have a large impact on the cavity polariton dispersion below the light line since we have seen that it greatly affects the spoof plasmon dispersion. Above the light line, we see a complex set of (bright region) cavity modes with increasing continuum density of states for larger mirror separations.

\section{Conclusions}
\label{Conclusions}

We have demonstrated that geometrically induced surface modes, or spoof plasmons, on metallic gratings result in cavity polariton modes in nanostructured cavities. 
These spoof surface plasmon modes strongly confine light to the grating surface and can be designed to have dispersion in the infrared and THz regions of the spectrum.  While these spoof surface modes are predicted for perfect conductive gratings, we have shown that these surface modes are also present on gratings with finite conductivity.  
By forming  a cavity from  a  periodic  grating and a planar mirror, cavity polariton modes are observed below the light-line and  exhibit dispersive  polaritonic resonances at short mirror separations.  We have derived an analytic expression for the modes of the perfectly conducting grating-plane cavity, and compared them with the numerically computed modes for a gold grating-plane cavity. These latter one were obtained using an RCWA approach. We have also introduced the differential density of states, whose zeros and poles contain information about the resonant modes of the grating-plane cavity problem as well as the individual surfaces. 
The differential density of states is a very useful indicator of the mode dispersion and hybridization in planar nanostructured cavities.  It provides insight into the origin of cavity resonances based on its sign.  Cavity modes (bright lines in  Figs. \ref{cavity_disp} and \ref{fig6}) move in the $k - k_x$ plane as a function of cavity separation, while the isolated modes (dark lines in  Figs. \ref{cavity_disp} and \ref{fig6})  do not change position.  
Below the light line, evanescent modes arise from singularities in the grating and plane reflection matrices in Eq. (\ref{DiffDOS1}) and their interaction with the other surface. In particular the nanostructuring generates low frequency spoof polariton dispersion relations which strongly depend on the grating geometrical parameters.
Due to a lack of symmetry in the cavity the grating's spoof plasmons do not strongly hybridize and spoof polariton cavity modes are only slightly shifted with respect to the isolated ones.  These modes would exist in a grating-grating cavity where symmetry dictates stronger hybridization and even and odd parity cavity modes, similar to hybridized metal-vacuum-metal plasmon seen in planar-planar cavities \cite{Intravaia07,Haakh10}.  
We find that the geometrically induced cavity surface polariton modes  from the nanostructured grating strongly affects the differential density of modes in the nanostructured cavity at short grating-plane separations and low frequency.
By increasing the cavity separation,  spoof plasmon cavity modes quickly overlap with their isolated counterparts reducing their contribution to the differential density of states. 
The low frequency enhancement of the differential density of states with respect to the plane-plane cavity can lead to increased electromagnetic forces and energy transfer at short separations in the former geometry. This effect can be further enhanced in grating-grating cavities with identical gratings.

Although we have presented the cavity polaritons in a reflection based cavity,  an analysis of a transmissive periodic nanostructured surface above a planar mirror or ground plane leads to similar resonant cavity modes.   These modes can be coupled to by free space planewaves and lead to thin perfect absorbing structures.  These perfect absorbers can have resonant modes in the infrared and have been observed in infrared frequency selective surfaces on a dielectric spacer above a ground plane \cite{Peters2010}. 
It is expected that engineering the spoof surface plasmon dispersion can lead to a new class of energy harvesting devices based on these new infrared absorbing structures.

\section*{Acknowledgments}

This research was  funded by Sandia's and Los Alamos' Laboratory Directed Research and Development (LDRD) program. Sandia is a multiprogram laboratory operated by Sandia Corporation, a Lockheed Martin Company, for the United States Department of Energy's National Nuclear Security Administration under contract DE-AC04-94AL85000. This work was carried out under the auspices of the NNSA of the U.S. DOE at LANL under Award No. DEAC52-06NA25396. F.I. acknowledges financial support from the European Union Marie Curie People program through the Career Integration Grant No. 631571.

\bibliographystyle{osajnl.bst}

\begin{thebibliography}{10}
\newcommand{\enquote}[1]{``#1''}

\bibitem{Castanie2012}
A.~Castani\'{e} and D.~Felbacq, \enquote{{Confined plasmonic modes in a
  nanocavity},} \oc  \textbf{285}, 3353-3357 (2012).

\bibitem{Kurokawa2007}
Y.~Kurokawa and H.~Miyazaki, \enquote{{Metal-insulator-metal plasmon
  nanocavities: Analysis of optical properties},} \prb
  \textbf{75}, 035411 (2007).

\bibitem{Sturman2007}
B.~Sturman, E.~Podivilov, and M.~Gorkunov, \enquote{{Eigenmodes for
  metal-dielectric light-transmitting nanostructures},} \prb
  \textbf{76}, 125104 (2007).

\bibitem{zhang2009slow}
J.~Zhang, L.~Cai, W.~Bai, Y.~Xu, and G.~Song, \enquote{Slow light at terahertz
  frequencies in surface plasmon polariton assisted grating waveguide,} J. Appl. Phys.
   \textbf{106}, 103715 (2009).

\bibitem{kats2011spoof}
M.~A. Kats, D.~Woolf, R.~Blanchard, N.~Yu, and F.~Capasso, \enquote{Spoof
  plasmon analogue of metal-insulator-metal waveguides,} \opex
  \textbf{19}, 14860-14870 (2011).

\bibitem{shen2013ultrathin}
X.~Shen and T.~J. Cui, \enquote{Ultrathin plasmonic metamaterial for spoof
  localized surface plasmons,} Laser Photonics Reviews \textbf{8}, 146-151
  (2013).

\bibitem{sun2013spoof}
G.~Sun, J.~B. Khurgin, and D.~P. Tsai, \enquote{Spoof plasmon waveguide enabled
  ultrathin room temperature THz GaN quantum cascade laser: a feasibility
  study,} \opex \textbf{21}, 28054-28061 (2013).

\bibitem{Pendry2004b}
J.~B. Pendry, L.~Mart\'{\i}n-Moreno, and F.~J. Garcia-Vidal,
  \enquote{{Mimicking surface plasmons with structured surfaces.}} Science (New
  York, N.Y.) \textbf{305}, 847-848 (2004).

\bibitem{Garcia-Vidal2005}
F.~J. Garcia-Vidal, L.~Mart\'{\i}n-Moreno, and J.~B. Pendry, \enquote{{Surfaces
  with holes in them: new plasmonic metamaterials},} J. of Optics A: Pure
  and Applied Optics \textbf{7}, S97-S101 (2005).


\bibitem{Yu10}
N. Yu {\it et~al.}, \enquote{{ Designer spoof surface plasmon structures collimate
  terahertz laser beams},} Nat. Mater. {\bf 9},  730-735  (2010).

\bibitem{Rusina10}
A. Rusina, M. Durach, and M. Stockman, \enquote{{ Theory of spoof plasmons in real
  metals},} Appl. Phys. A {\bf 100},  375  (2010).


\bibitem{Peters2010}
D.~W. Peters, P.~Davids, J.~R. Wendt, A.~A. Cruz-Cabrera, S.~A. Kemme, and
  S.~Samora, \enquote{Metamaterial-inspired high-absorption surfaces for
  thermal infrared applications,} \pspie \textbf{7609}, 76091C
  (2010).

\bibitem{Peters2012}
D.~W. Peters, C.~M. Reinke, P.~S. Davids, J.~F. Klem, D.~Leonhardt, J.~R.
  Wendt, J.~K. Kim, and S.~Samora, \enquote{{Nanoantenna-enabled midwave
  infrared focal plane arrays},} \pspie \textbf{8353}, 83533B
  (2012).

\bibitem{Lambrecht2008}
A.~Lambrecht and V.~Marachevsky, \enquote{{Casimir interaction of dielectric
gratings},} \prl  \textbf{101}, 160403 (2008).

\bibitem{bimonte2009scattering}
G.~Bimonte, \enquote{Scattering approach to casimir forces and radiative heat
  transfer for nanostructured surfaces out of thermal equilibrium,} \pra \textbf{80}, 042102 (2009).

\bibitem{Davids2010b}
P.~S. Davids, F.~Intravaia, F.~D. S.~S. Rosa, and D.~Dalvit, \enquote{{Modal
  approach to Casimir forces in periodic structures},} \pra
  \textbf{82}, 062111 (2010).

\bibitem{DiegoHeatTransfer}
R.~Gu\'erout, J.~Lussange, F.~S.~S. Rosa, J.-P. Hugonin, D.~A.~R. Dalvit, J.-J.
  Greffet, A.~Lambrecht, and S.~Reynaud, \enquote{Enhanced radiative heat
  transfer between nanostructured gold plates,} \prb \textbf{85},
  180301 (2012).

\bibitem{Lussange2012}
J.~Lussange, R.~Gu\'{e}rout, F.~S.~S. Rosa, J.-J. Greffet, A.~Lambrecht, and
  S.~Reynaud, \enquote{{Radiative heat transfer between two dielectric
  nanogratings in the scattering approach},} \prb \textbf{86},
  085432 (2012).

\bibitem{Reid2013}
M.~T.~H. Reid, J.~White, and S.~G. Johnson, \enquote{{Fluctuating surface
  currents: An algorithm for efficient prediction of Casimir interactions among
  arbitrary materials in arbitrary geometries},} \pra  \textbf{88},
  022514 (2013).

\bibitem{rodriguez2011casimir}
A.~W. Rodriguez, F.~Capasso, and S.~G. Johnson, \enquote{The Casimir effect in
  microstructured geometries,} Nature Photonics \textbf{5}, 211-221 (2011).

\bibitem{Intravaia2013}
F.~Intravaia, S.~Koev, I.~W. Jung,A.~A. Talin, P.~S. Davids, R.~S. Decca,
  V.~a. Aksyuk, D.~A.~R. Dalvit, and D.~L\'{o}pez, \enquote{{Strong Casimir
  force reduction through metallic surface nanostructuring.}} Nature
  Communications \textbf{4}, 3515 (2013).

\bibitem{Polder71}
D. Polder and M. Van~Hove, \enquote{Theory of radiative heat transfer between
  closely spaced bodies}, \prb {\bf 4},  3303-3314  (1971).

\bibitem{Carminati1999}
R.~Carminati and J.-J. Greffet, \enquote{Near-field effects in spatial
  coherence of thermal sources,} \prl  \textbf{82}, 1660-1663
  (1999).

\bibitem{Narayanaswamy2008}
A.~Narayanaswamy, S.~Shen, and G.~Chen, \enquote{{Near-field radiative heat
  transfer between a sphere and a substrate},} \prb \textbf{78},
  115303 (2008).

\bibitem{Narayanaswamy2009}
A.~Narayanaswamy, S.~Shen, L.~Hu, X.~Chen, and G.~Chen, \enquote{{Breakdown of
  the Planck blackbody radiation law at nanoscale gaps},} Appl. Phys. A
  \textbf{96}, 357-362 (2009).

\bibitem{Ottens2011}
R.~Ottens, V.~Quetschke, S.~Wise, A.~Alemi, R.~Lundock, G.~Mueller, D.~Reitze,
  D.~Tanner, and B.~Whiting, \enquote{{Near-Field Radiative Heat Transfer
  between Macroscopic Planar Surfaces},} \prl \textbf{107},
  014301 (2011).

\bibitem{Biehs2011}
S.-A. Biehs, F.~S.~S. Rosa, and P.~Ben-Abdallah, \enquote{{Modulation of
  near-field heat transfer between two gratings},} \apl
  \textbf{98}, 243102 (2011).

\bibitem{Jiang2009}
T.~Jiang, L.~Shen, X.~Zhang, and L.~Ran, \enquote{{High-order modes of spoof
  surface plasmon polaritons on periodically corrugated metal surfaces},}
  Progress In Electromagnetics Research \textbf{8}, 91-96 (2009).

\bibitem{intravaia2012quasianalytical}
F.~Intravaia, P.~Davids, R.~Decca, V.~Aksyuk, D.~L{\'o}pez, and D.~Dalvit,
  \enquote{Quasianalytical modal approach for computing casimir interactions in
  periodic nanostructures,} \pra  \textbf{86}, 042101 (2012).

\bibitem{Li1996}
L.~Li, \enquote{{Use of Fourier series in the analysis of discontinuous
  periodic structures},} \josaa  \textbf{13}, 1870-1876 (1996).

\bibitem{Li1996a}
L.~Li, \enquote{{Formulation and comparison of two recursive matrix algorithms
  for modeling layered diffraction gratings},} \josaa \textbf{13}, 1024-1035
  (1996).

\bibitem{Li1997}
L.~Li, \enquote{{New formulation of the Fourier modal method for crossed
  surface-relief gratings},} \josaa 
  \textbf{14}, 2758-2767 (1997).

\bibitem{Li2003}
Z.-Y. Li and K.-M. Ho, \enquote{{Analytic modal solution to light propagation
  through layer-by-layer metallic photonic crystals},} \prb
  \textbf{67}, 165104 (2003).

\bibitem{Li2003c}
Z.-Y. Li and L.-L. Lin, \enquote{{Photonic band structures solved by a
  plane-wave-based transfer-matrix method},} \pre \textbf{67},
  046607 (2003).

\bibitem{wirzba2008casimir}
A.~Wirzba, \enquote{The Casimir effect as a scattering problem,} J. of
  Physics A: Mathematical and Theoretical \textbf{41}, 164003 (2008).

\bibitem{Intravaia07}
F. Intravaia, C. Henkel, and A. Lambrecht, \enquote{Role of surface plasmons in the
  Casimir effect,} \pra {\bf 76},  033820  (2007).

\bibitem{Haakh10}
H. Haakh, F. Intravaia, and C. Henkel,  \enquote{Temperature dependence of the
  plasmonic Casimir interaction}, \pra {\bf 82},  012507  (2010).



\end{thebibliography}

\end{document}